\documentclass[12pt,letterpaper]{article}

\usepackage[utf8]{inputenc}
\usepackage[english]{babel}
\usepackage{amsmath}
\usepackage{amsfonts}
\usepackage{amssymb}
\usepackage{graphicx}
\usepackage[left=25.4mm,right=24.2mm,top=25.4mm,bottom=25.4mm]{geometry} 
\usepackage{changepage} 
\usepackage{times} 
\usepackage{titlesec} 
\usepackage{cite} 
\usepackage[colorlinks=true,citecolor=cyan,linkcolor=red,bookmarks=true,bookmarksnumbered=true]{hyperref} 
\usepackage{tikz} 
\usetikzlibrary{arrows.meta} 
\usepackage[font=footnotesize,labelfont=bf]{caption} 
\usepackage{subfigure}

\renewcommand{\thesection}{\Roman{section}}
\renewcommand{\thesubsection}{\arabic{subsection}}

\titleformat{\section}[hang]{\bf\large}{\thesection\enspace}{5pt}{}
\titleformat{\subsection}[hang]{\bf\normalsize}{\thesubsection\enspace}{5pt}{}

\begin{document}

{
\flushleft\Huge\bf Supersymmetric Quantum Mechanics, multiphoton algebras and coherent states 
}\vspace{5mm}

\begin{adjustwidth}{0.75in}{} 
{
\flushleft\bf Juan D Garc{\'i}a-Mu{\~n}oz$^{1,2}$\footnote{Author to whom correspondence should be addressed}, David J Fern{\'a}ndez C$^1$ and F Vergara-M{\'e}ndez$^{1,3}$
}
\vspace{2.5mm}
{
\par\noindent\small $^1$Physics Department, Cinvestav, P.O.B. 14-740, 07000 Mexico City, Mexico \\
$^2$Instituto de F{\'i}sica y Matem{\'a}ticas, Universidad Michoacana de San Nicolás de Hidalgo, Edificio C-3, Ciudad Universitaria, Francisco J. M{\'u}jica S/N Col. Fel{\'i}citas del R{\'i}o, 58040 Morelia, Michoac{\'a}n, M{\'e}xico.\\
$^3$Instituto Thomas Jefferson Campus Santa M{\'o}nica, Gardenia 5, Ex-hacienda de Santa Monica, 54050 Tlanepantla de Baz, Estado de M{\'e}xico, M{\'e}xico.
}
\vspace{2.5mm}
{
\par\noindent\footnotesize Email: juan.domingo.garcia@umich.mx, david.fernandez@cinvestav.mx and fer335@gmail.com
}
\vspace{5mm}
{
\par\noindent\bf Abstract
}
{ 
\newline\small  The multiphoton algebras for one-dimensional Hamiltonians with infinite discrete spectrum, and for their associated $k$th-order SUSY partners are studied. In both cases, such an algebra is generated by the multiphoton annihilation and creation operators, as well as by Hamiltonians which are functions of an appropriate number operator. The algebras obtained turn out to be polynomial deformations of the corresponding single-photon algebra previously studied. The Barut-Girardello coherent states, which are eigenstates of the annihilation operator, are obtained and their uncertainty relations are explored by means of the associated quadratures.  
}
\vspace{2mm}
{
\newline\footnotesize {\bf Keywords:} Supersymmetric Quantum Mechanics, multiphoton algebras, coherent states 
}
\end{adjustwidth}

\section{Introduction}

Supersymmetric Quantum Mechanics (SUSY-QM) was introduced in 1981 by Witten as a toy model to understand supersymmetry, in which the involved bosons and fermions are connected to each other \cite{Witten1981}. It is a useful tool to study exactly solvable one-dimensional potentials in Quantum Mechanics \cite{Cooper1995}. It is closely related as well with other approaches used previously to address quantum mechanical eigenproblems, as the factorization method, intertwining technique and Darboux transformation \cite{Fernandez2019}. Concerning Darboux transformations, they have been employed in soliton theory for generating new solutions (potentials) departing from a given initial one, where they are called B{\"a}cklund-Darboux transformations (see e.g. \cite{Matveev1991}). On the other hand, the factorization method was employed at first to identify the family of potentials whose analytic solution can be generated in a closed form. This was the key idea used by Dirac in 1935 to solve the harmonic oscillator potential \cite{Dirac1935}. Later on, Schr{\"o}dinger realized that the technique could be applied as well to the hydrogen atom \cite{Schrodinger1940,Schrodinger1941}. Then, Infeld and his collaborators expanded the set of potentials that can be solved using such a technique \cite{Infeld1941,Infeld1945}, publishing in 1951 a seminal article containing a classification of the families of potentials that can be solved through factorization \cite{Infeld1951}. This article became so influential among the physicist that, for a time, it was thought that the factorization method was essentially exhausted. However, in 1984 Mielnik introduced a generalization of the factorization method which represented a breakthrough on the subject. In fact, he showed that the standard annihilation and creation operators factorizing the harmonic oscillator Hamiltonian in a given order are not unique \cite{Bogdan1984}. Moreover, if the  initial product of the generalized factorization operators is permuted, then a new Hamiltonian appears, which has the same spectrum as the harmonic oscillator but different eigenstates. Later on, another authors generalized the idea raised by Mielnik in several ways, thus generating new solvable Hamiltonians through the factorization method. Shortly after, it was discovered its connection with SUSY-QM, in which a SUSY partner Hamiltonian is generated from a given initial one. These SUSY partner Hamiltonians have spectra slightly different from the initial one, but sharing an isospectral part. Among the vast literature that exists about SUSY-QM we recommend \cite{Cooper1995,Junker1996,bagchi2000,Mielnik2004,Fernandez2010,Fernandez2019,Junker1998,Nico2004,Alonso2008,Gangopadhyaya2018,Andrianov2007,Andrianov2012}.   

On the other hand, the SUSY partner Hamiltonians of the harmonic oscillator besides the corresponding annihilation and creation operators, generate what nowadays are called polynomial Heisenberg algebras \cite{Veronique1999,Carballo2004}. It has been shown as well that the harmonic oscillator Hamiltonian, together with annihilation and creation operators that are the $m$th powers of $a$ and $a^{\dagger}$ (usually called multiphoton operators) also generate a polynomial Heisenberg algebra \cite{Celeita2016,Celeita2019}. In addition, if for the SUSY partners of the harmonic oscillator Hamiltonian the multiphoton annihilation and creation operators are employed, then additional polynomial Heisenberg algebras can be generated. Let us note that the harmonic oscillator Hamiltonian and its SUSY partners are almost the only systems for which the corresponding algebras have been explored \cite{Veronique1999,Carballo2004,Marquette2013,Marquette2014,Gómez-Ullate2014,MOrales2022}. Thus, it is natural to ask ourselves if it is possible to implement a similar approach for other one dimensional potentials (see e.g. \cite{Veronique2007}). Another natural issue is to figure out which algebras rule these systems, and their corresponding SUSY partners, if multiphoton annihilation and creation operators are used. 

In this paper, we will try to answer these questions. In order to do that, the article has been organized as follows. In section II, it is sketched how to build the $k$th-order SUSY partners $H_{k}$ departing from the initial Hamiltonian $H_{0}$, which has an infinite discrete spectrum. Then, in section III, we will explore the multiphoton algebra associated to these Hamiltonians together with the corresponding multiphoton annihilation and creation operators. In section IV, we will generate the Barut-Girardello coherent states for this algebra, and we will explore the uncertainty relation for the natural quadratures. Then, in section V, we will focus on the harmonic oscillator and trigonometric P{\"o}schl-Teller potentials as particular cases. Finally, in section VI we close the paper with our conclusions.

\section{$\mathbf{k}$th-order supersymmetric quantum mechanics}

\begin{figure}[th]
\begin{center}
\begin{tikzpicture}

\draw (0,0)--(2,0) (0,2)--(2,2) (0,2.5)--(2,2.5) (0,3.5)--(2,3.5) (0,4.5)--(2,4.5) (0,6.5)--(2,6.5) (6,2.5)--(8,2.5) (6,3.5)--(8,3.5) (6,4.5)--(8,4.5) (6,6.5)--(8,6.5);

\fill (1,0.7) circle [radius=1.5pt] (1,1.1) circle [radius=1.5pt] (1,1.5) circle [radius=1.5pt] (1,5.2) circle [radius=1.5pt] (1,5.6) circle [radius=1.5pt] (1,6) circle [radius=1.5pt] (1,7.2) circle [radius=1.5pt] (1,7.6) circle [radius=1.5pt] (1,8) circle [radius=1.5pt] (7,5.2) circle [radius=1.5pt] (7,5.6) circle [radius=1.5pt] (7,6) circle [radius=1.5pt] (7,7.2) circle [radius=1.5pt] (7,7.6) circle [radius=1.5pt] (7,8) circle [radius=1.5pt];

\draw[arrows={-Stealth[scale=1.5]}] (2.5,2.5)--(5.5,2.5);
\draw[arrows={Stealth[scale=1.5]-}] (2.5,3.5)--(5.5,3.5);
\draw[arrows={-Stealth[scale=1.5]}] (2.5,4.5)--(5.5,4.5);
\draw[arrows={-Stealth[scale=1.5]}] (5.5,4.5)--(5.5,3.5);
\draw[arrows={-Stealth[scale=1.5]}] (5.5,2.5)--(5.5,3.5);

\draw (-0.5,0) node{\large $\epsilon_{k}$} (-0.5,2) node{\large $\epsilon_{1}$} (-0.5,2.5) node{\large $E_{0}$} (-0.5,3.5) node{\large $E_{1}$} (-0.5,4.5) node{\large $E_{2}$} (-0.5,6.5) node{\large $E_{n}$} (8.5,2.5) node{\large $E_{0}$} (8.5,3.5) node{\large $E_{1}$} (8.5,4.5) node{\large $E_{2}$} (8.5,6.5) node{\large $E_{n}$} (4,2.5) node[above]{\large $B_{k}$} (4,3.5) node[above]{\large $B^{\dagger}_{k}$} (4,4.5) node[above]{\large $B_{k}$} (5.5,4) node[right]{\large $a$} (5.5,3) node[right]{\large $a^{\dagger}$} (1,8.5) node[above]{\Large $H_{k}$} (7,8.5) node[above]{\Large $H_{0}$};
\end{tikzpicture} 
\caption{Spectral representation of $H_{0}$, $H_{k}$ and the intertwinings linking them.} \label{F-1}
\end{center}
\end{figure}
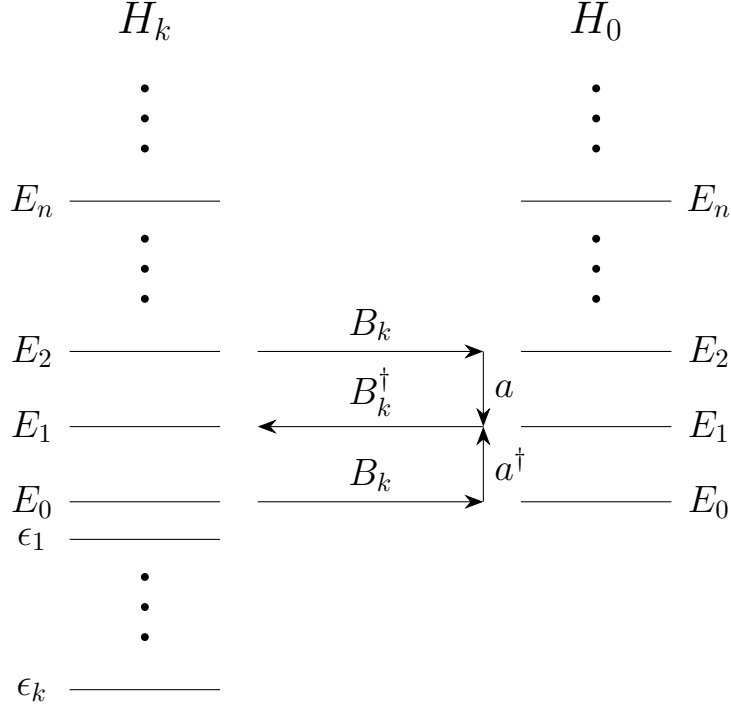

Let us consider a chain of one-dimensional Hamiltonians of the form
\begin{equation} \label{E-1}
H_{j} = -\frac{1}{2}\frac{d^{2}}{dx^{2}} + V_{j}(x),\quad j = 0,1,\dots,k,
\end{equation}
with $k$ being a natural number. It is said that the Hamiltonian $H_{k}$ is a $k$th-order supersymmetric partner of the initial one $H_{0}$ if both are intertwined as follows 
\begin{equation} \label{E-2}
H_{k}B^{\dagger}_{k} = B^{\dagger}_{k}H_{0},
\end{equation}
where $B_{k}^{\dagger}$ is a $k$th-order differential intertwining operator which is factorized in the way
\begin{equation} \label{E-3}
B_{k}^{\dagger} = b_{k}^{\dagger}\cdots b_{1}^{\dagger},
\end{equation}  
with $b_{j}^{\dagger}$ being the first-order differential operator
\begin{equation} \label{E-4}
b_{j}^{\dagger}  = \frac{1}{\sqrt{2}}\left(-\frac{d}{dx} + \beta_{j}\left(x, \epsilon_{j}\right)\right),
\end{equation}
whose adjoint operator is 
\begin{equation} \label{E-5}
b_{j} = \frac{1}{\sqrt{2}}\left(\frac{d}{dx} + \beta_{j}\left(x, \epsilon_{j}\right)\right),
\end{equation}
where the functions $\beta_{j}(x, \epsilon_{j})$ are real. The quantities $\epsilon_{j}$ are called factorization energies, which for simplicity will be taken as $\epsilon_{k} < \dots < \epsilon_{2} < \epsilon_{1} < E_{0}$, with $E_{0}$ denoting the ground state energy of $H_{0}$ \cite{Nico2004}. Each factorization energy $\epsilon_{j}$ is associated to one operator $b_{j}^{\dagger}$, which satisfies an intertwining relation similar to Eq.~\eqref{E-2} but involving $H_{j}$ and $H_{j-1}$, i.e., the Hamiltonian $H_{k}$ is the result of an iterative process leading to   
\begin{equation} \label{E-6}
H_{0} = b_{1}b_{1}^{\dagger} + \epsilon_{1},\ H_{j} = b_{j}^{\dagger}b_{j} + \epsilon_{j} = b_{j+1}b_{j+1}^{\dagger} + \epsilon_{j+1},\ H_{k} = b_{k}^{\dagger}b_{k} + \epsilon_{k},\ j=1,2,\dots,k-1.  
\end{equation} 
Let us suppose that the Hamiltonian $H_{0}$ is solvable, i.e., we know its eigenvalues $E_{n}$ and eigenfunctions $\psi^{(0)}_{n}$. From the intertwining relation \eqref{E-2} we realize that the Hamiltonian $H_{k}$ has also $E_{n}$ as its eigenvalue, while the corresponding eigenfunctions $\psi^{(k)}_{n}$ and $\psi^{(0)}_{n}$ are related as follows
\begin{equation} \label{E-7}
\psi^{(k)}_{n} = \frac{B^{\dagger}_{k}\psi^{(0)}_{n}}{\sqrt{(E_{n} - \epsilon_{1})\dots (E_{n} - \epsilon_{k})}},\quad \psi^{(0)}_{n} = \frac{B_{k}\psi^{(k)}_{n}}{\sqrt{(E_{n} - \epsilon_{1})\dots (E_{n} - \epsilon_{k})}}.
\end{equation} 
Furthermore, the products of the operator $B^{\dagger}_{k}$ and its adjoint $B_{k}$ turn out to be 
\begin{equation} \label{E-9}
B^{\dagger}_{k}B_{k} = (H_{k} - \epsilon_{1})\dots (H_{k} - \epsilon_{k}),\quad 
B_{k}B^{\dagger}_{k} = (H_{0} - \epsilon_{1})\dots (H_{0} - \epsilon_{k}).
\end{equation}
It is worth noticing that there exist $k$ functions $\psi^{(k)}_{\epsilon_{j}}$ that are in the {\it kernel} of the operators $B_{k}$ and $B_{k}^{\dagger}B_{k}$, i.e., $\psi_{\epsilon_{j}}^{(k)}$ is a solution of the eigenvalue equation for the Hamiltonian $H_{k}$ associated to the factorization energy $\epsilon_{j}$. If such solutions are square-integrable, the factorization energies $\epsilon_{j}$ must lie in the spectrum of the $k$th-order SUSY partner Hamiltonian $H_{k}$, thus $\mathrm{Sp}(H_{k}) = \{\epsilon_{k},\dots,\epsilon_{1},E_{n}, n = 0,1,2,\dots\}$. In Fig. \ref{F-1} it is sketched the energy levels of both Hamiltonians $H_{0}$ and $H_{k}$ as well as the intertwinings linking them.  
    	
On the other hand, the functions $\beta_{j}\left(x, \epsilon_{j}\right)$ fulfill the finite difference formula
\begin{equation} \label{E-10}
\beta_{j}\left(x, \epsilon_{j}\right) = -\beta_{j-1}\left(x,\epsilon_{j-1}\right) - \frac{2\left(\epsilon_{j - 1} - \epsilon_{j}\right)}{\beta_{j-1}\left(x,\epsilon_{j - 1}\right) - \beta_{j - 1}\left(x,\epsilon_{j}\right)},\quad j = 2,3,\dots,k.
\end{equation}
Thus, we only need $k$ initial functions $\beta_{1}\left(x,\epsilon_{j}\right)$ of the finite difference equation \eqref{E-10}, which are as well solutions of the initial Riccati equation
\begin{equation} \label{E-11}
\beta'_{1}\left(x,\epsilon_{j}\right) + \beta^{2}_{1}\left(x,\epsilon_{j}\right) = 2\left(V_{0} - \epsilon_{j}\right),
\end{equation} 
where $f'(x)\equiv df(x)/dx$. If we make the change of variable $\beta_{1}\left(x,\epsilon_{j}\right) = u'_{j}/u_{j}$, the previous equation can be written in the most familiar form 
\begin{equation} \label{E-12}
-\frac{1}{2}u''_{j} + V_{0}(x)u_{j} = \epsilon_{j}u_{j},
\end{equation}
i.e., $u_{j}\equiv u(x,\epsilon_{j})$ is a solution of the stationary Schr{\"o}dinger equation for the initial Hamiltonian $H_{0}$. Meanwhile, the potential $V_{k}(x)$ is given by
\begin{equation} \label{E-13}
V_{k}(x) = V_{0} - \sum\limits_{j=1}^{k}\beta'_{j}\left(x,\epsilon_{j}\right) = V_{0} - \left\{\log\left[W\left(u_{1},\dots,u_{k}\right)\right]\right\}''.
\end{equation} 

\section{Multiphoton algebras of $\mathbf{H_{0}}$ and $\mathbf{H_{k}}$}
In order to address the multiphoton algebras of $H_{0}$ and $H_{k}$ it is important to study first the intrinsic algebraic structure of $H_{0}$. 

\subsection{Intrinsic algebra of $\mathbf{H_{0}}$}

We are going to consider one-dimensional Hamiltonians with the form given in Eq.~\eqref{E-1} for $j = 0$. We suppose that $H_{0}$ is solvable, i.e., its eigenvalues $E_{n}$ and eigenstates $|\psi_{n}^{(0)}\rangle$ are known. Moreover, the eigenvalues $E_{n}$ are ordered in the standard way, and define an explicit dependence on the principal quantum number $n$, i.e., $E_{n}\equiv E(n)$.

The intrinsic annihilation and creation operators $a$ and $a^{\dagger}$ are defined through their action onto the normalized states $|\psi_{n}^{(0)}\rangle$ as follows
\begin{equation} \label{E-14}
a|\psi_{n}^{(0)}\rangle = r(n)|\psi_{n-1}^{(0)}\rangle,\ a^{\dagger}|\psi_{n}^{(0)}\rangle = \overline{r}(n+1)|\psi_{n+1}^{(0)}\rangle,\ r(n) = e^{i\alpha(E_{n}-E_{n-1})}\sqrt{E_{n}-E_{0}},
\end{equation}    
where $\overline{z}$ denotes the complex conjugate of $z\in\mathbb{C}$ and $\alpha\in\mathbb{R}$ is a free phase parameter. Thus, the set of operators $\{H_{0}, a, a^{\dagger}\}$ generates what is called intrinsic algebra for the system, which is defined through the following commutators
\begin{equation} \label{E-16}
[a, a^{\dagger}] = f(N),\ [H_{0}, a^{\dagger}] = f(N-1)a^{\dagger},\ [H_{0}, a] = -f(N)a,\ f(N) = E(N+1) - E(N),
\end{equation}
with $f(N)$ being a function of the intrinsic number operator $N$ which fulfills 
\begin{equation} \label{E-18}
N|\psi_{n}^{(0)}\rangle = n|\psi_{n}^{(0)}\rangle,\quad [N, a^{\dagger}] = a^{\dagger},\quad [N, a] = -a. 
\end{equation}

\subsection{Multiphoton algebra of $\mathbf{H_{0}}$}

\begin{figure}[ht]
\begin{center}
\begin{tikzpicture}

\draw (0,0) node[right]{$j=1$} (3.3,0) node[right]{$j=0$} (-5,0) node[right]{$j=m-1$} (0.65,6.5) node[above]{\large $H_{0}$} (7,6.5) node[above]{\large $H_{k}$};

\draw (6.3,0) node[right]{$\epsilon_{1},\ \left|\psi_{\epsilon_{1}}^{(k)}\right\rangle$} (6.3,-2) node[right]{$\epsilon_{k},\ \left|\psi_{\epsilon_{k}}^{(k)}\right\rangle$};

\draw (3,0.75) node[right]{$E_{0},\ |\psi_{0}^{(0)}\rangle$} (3,3.5) node[right]{$E_{m},\ |\psi_{m}^{(0)}\rangle$};

\draw (-0.3,1.5) node[right]{$E_{1},\ |\psi_{1}^{(0)}\rangle$} (-0.75,4.25) node[right]{$E_{m+1},\ |\psi_{m+1}^{(0)}\rangle$};	

\draw (-5.3,2.75) node[right]{$E_{m-1},\ |\psi_{m-1}^{(0)}\rangle$} (-5.3,5.5) node[right]{$E_{2m-1},\ |\psi_{2m-1}^{(0)}\rangle$};

\fill (-2,0) circle[radius=1.25pt] (-1.5,0) circle[radius=1.25pt] (-1,0) circle[radius=1.25pt] (4,4) circle[radius=1.25pt] (4,4.25) circle[radius=1.25pt] (4,4.5) circle[radius=1.25pt] (0.65,4.75) circle[radius=1.25pt] (0.65,5) circle[radius=1.25pt] (0.65,5.25) circle[radius=1.25pt] (-3.75,6) circle[radius=1.25pt] (-3.75,6.25) circle[radius=1.25pt] (-3.75,6.5) circle[radius=1.25pt] (-1,1.75) circle[radius=1.25pt] (-1.5,2) circle[radius=1.25pt] (-2,2.25) circle[radius=1.25pt] (-1,4.5) circle[radius=1.25pt] (-1.5,4.75) circle[radius=1.25pt] (-2,5) circle[radius=1.25pt] (7,-0.75) circle[radius=1.25pt] (7,-1) circle[radius=1.25pt] (7,-1.25) circle[radius=1.25pt];

\draw[loosely dashed,color=gray] (5.65,7)--(5.65,-2.5); 
\end{tikzpicture}
\caption{Spectrum of $H_{0}$, divided into $m$ infinite ladders labeled by the index $j$. The spectrum of $H_{k}$ has a similar splitting as $\mathrm{Sp}(H_{0})$, but it contains $k$ additional single-step ladders corresponding to the factorization energies $\epsilon_{i}$.} \label{F-2}
\end{center}
\end{figure}
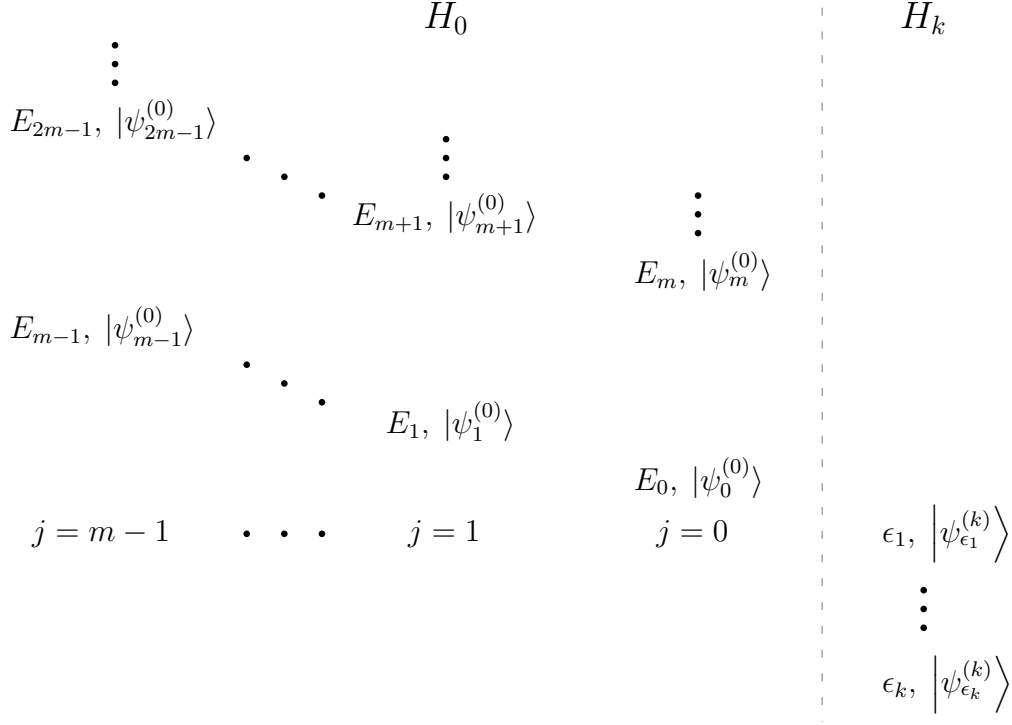

Let us consider now the multiphoton annihilation and creation operators $a_{m}, a_{m}^{\dagger}$ defined as the $m$th powers of the intrinsic ones
\begin{equation} \label{E-19}
a_{m} = a^{m},\quad a^{\dagger}_{m} = (a^{\dagger})^{m},\quad m\in\mathbb{N},
\end{equation}
whose actions onto the normalized eigenstates $|\psi_{n}^{(0)}\rangle$ are given by
\begin{equation} \label{E-20}
\begin{aligned}
a_{m}|\psi_{n}^{(0)}\rangle &= e^{i\alpha(E_{n}-E_{n-m})}\sqrt{(E_{n}-E_{0})\dots (E_{n-m+1}-E_{0})}|\psi_{n-m}^{(0)}\rangle, \\
a_{m}^{\dagger}|\psi_{n}^{(0)}\rangle &= e^{-i\alpha(E_{n+m}-E_{n})}\sqrt{(E_{n+1}-E_{0})\dots (E_{n+m}-E_{0})}|\psi_{n+m}^{(0)}\rangle,
\end{aligned}
\end{equation}
with $\alpha\in\mathbb{R}$ being again a free phase parameter. Analogously to the intrinsic case, the multiphoton algebra is defined by the commutation rules
\begin{equation} \label{E-21}
\begin{aligned}
&[a_{m},a_{m}^{\dagger}] = \prod_{l=0}^{m-1}\left[E(N+m-l)-E_{0}\right] - \prod_{l=0}^{m-1}\left[E(N-l)-E_{0}\right], \\
&[H_{0},a^{\dagger}_{m}] = f_{m}(N-m)a^{\dagger}_{m},\quad [H_{0},a_{m}] = -f_{m}(N)a_{m},\quad f_{m}(N) = E(N+m)-E(N).
\end{aligned}
\end{equation}
It is important to define now the extreme states, which are eigenstates of the Hamiltonian $H_{0}$ that are also annihilated by the multiphoton annihilation operator, i.e., $a_{m}|\psi_{j}^{(0)}\rangle=0$. The intrinsic algebra of $H_{0}$ is the particular case of the multiphoton algebra with $m=1$, for which there exists only one extreme state, namely, the ground state $|\psi_{0}^{(0)}\rangle$. The iterated action of the intrinsic creation operator on such extreme state leads to the infinite ladder of eigenstates $|\psi_{n}^{(0)}\rangle$ of $H_{0}$. On the other hand, in order to identify the extreme states $|\psi_{j}^{(0)}\rangle$ for general $m>1$, let us note that they are also in the \textit{kernel} of the operator $a_{m}^{\dagger}a_{m}$, i.e.,
\begin{equation} \label{E-23}
\begin{aligned}
a_{m}^{\dagger}a_{m}|\psi_{j}^{(0)}\rangle=\prod\limits_{l=0}^{m-1}[E(j-l)-E_0]|\psi_{j}^{(0)}\rangle=(E_j-E_0)(E_{j-1}-E_0)\cdots (E_{j-m+1}-E_0)|\psi_{j}^{(0)}\rangle=0.
\end{aligned}
\end{equation}
We can see that the eigenstates $|\psi_{j}^{(0)}\rangle$ of $H_{0}$ with $j=0,1,...,m-1$ are as well annihilated by $a_{m}$, thus, the extreme states are the first $m$ energy eigenstates of $H_{0}$. Moreover, the iterated action of the multiphoton creation operator on each extreme state $|\psi_{j}^{(0)}\rangle$ produces subsequent states, denoted as $|\psi_{j+nm}^{(0)}\rangle$, whose energies $E_{j+nm}$ at the end arranges as $j$ infinite ladders. We note that such energy ladders do not only split the spectrum of the system, but also divide the state space in $m$ orthogonal subspaces, such that their direct sum is the original space. In Fig. \ref{F-2} it is represented the spectrum of $H_{0}$, divided in $m$ infinite independent ladders labeled by the index $j$. Let us note as well that the energy levels of $H_{0}$ do not have to be equidistant, although in such figure we are showing them in this way for illustration purposes.

\subsection{Multiphoton algebra of $\mathbf{H_{k}}$}

For the $k$th SUSY partner Hamiltonian $H_{k}$ the natural multiphoton annihilation and creation operators are given by
\begin{equation}\label{E-24}
\ell_{k,m} = B^{\dagger}_{k}a_{m}B_{k},\quad \ell^{\dagger}_{k,m} = B^{\dagger}_{k}a^{\dagger}_{m}B_{k},
\end{equation}
where $B_{k}$, $B_{k}^{\dagger}$ are the intertwining operators of the previous section, while $a_{m}$, $a_{m}^{\dagger}$ are the multiphoton annihilation and creation operators for the initial Hamiltonian $H_{0}$. Of course, it is possible to take into account another multiphoton annihilation and creation operators, e.g., to choose $\left(\ell_{k,1}\right)^{m}$ and $\left(\ell_{k,1}^{\dagger}\right)^{m}$. However, the last turn out to be polynomial deformations of the ones given in \eqref{E-24}; this is the reason why from now on we will just consider $\ell_{k,m}$ and $\ell_{k,m}^{\dagger}$. We can find the action of $\ell_{k,m}$, $\ell_{k,m}^{\dagger}$ onto the states $\left|\psi_{n}^{(k)}\right\rangle$ by using Eqs.~(\ref{E-7},~\ref{E-20}), leading to
\begin{equation} \label{E-25}
\begin{aligned}
\ell_{k,m}\left|\psi_{n}^{(k)}\right\rangle &= e^{i\alpha(E_{n}-E_{n-m})}\sqrt{\prod_{l=0}^{m-1}(E_{n-l}-E_{0})\prod_{i=1}^{k}(E_{n}-\epsilon_{i})(E_{n-m}-\epsilon_{i})}\left|\psi_{n-m}^{(k)}\right\rangle, \\
\ell_{k,m}^{\dagger}\left|\psi_{n}^{(k)}\right\rangle &= e^{-i\alpha(E_{n+m}-E_{n})}\sqrt{\prod_{l=0}^{m-1}(E_{n+m-l}-E_{0})\prod_{i=1}^{k}(E_{n}-\epsilon_{i})(E_{n+m}-\epsilon_{i})}\left|\psi_{n+m}^{(k)}\right\rangle,
\end{aligned}
\end{equation}
where $\alpha\in\mathbb{R}$ is a free phase parameter. Before exploring the multiphoton algebra generated by the operator set $\{H_{k},\ell_{k,m},\ell_{k,m}^{\dagger}\}$, let us identify the number operator $N_{k}$, such that its action onto the states $\left|\psi_{n}^{(k)}\right\rangle$ is
\begin{equation} \label{E-26}
N_{k}\left|\psi_{n}^{(k)}\right\rangle = n\left|\psi_{n}^{(k)}\right\rangle.
\end{equation}
This operator can be written in terms of the intrinsic number operator $N$ of the initial system defined in equation \eqref{E-18}, whose explicit expression becomes
\begin{equation} \label{E-27}
N_{k} = B_{k}^{\dagger}\frac{N}{(E(N)-\epsilon_{1})\dots(E(N)-\epsilon_{k})}B_{k}.
\end{equation}
Since the states $\left|\psi_{\epsilon_{i}}^{(k)}\right\rangle$ are in the {\it kernel} of $B_{k}$, then $N_{k}\left|\psi_{\epsilon_{i}}^{(k)}\right\rangle = 0$, due to they are not characterized by the quantum number $n$ but by the factorization energy $\epsilon_{i}$.

By using Eq.~\eqref{E-25}, the multiphoton algebra generated by $\{H_{k},\ell_{k,m},\ell_{k,m}^{\dagger}\}$ is defined by the commuta{\-t}ion rules 
\begin{align} 
&\begin{aligned}
&[\ell_{k,m},\ell_{k,m}^{\dagger}] = \Bigg\{\prod_{l=0}^{m-1}[E(N_{k}+m-l)-E_{0}]\prod_{i=1}^{k}[E(N_{k}+m)-\epsilon_{i}][E(N_{k})-\epsilon_{i}]\\
&\qquad\qquad\ \ -\prod_{l=0}^{m-1}[E(N_{k}-l)-E_{0}]\prod_{i=1}^{k}[E(N_{k}-m)-\epsilon_{i}][E(N_{k})-\epsilon_{i}]\Bigg\}\mathcal{P}^{(k)}_{iso}, \label{E-28}
\end{aligned} \\
&[H_{k},\ell_{k,m}^{\dagger}] = f_{m}(N_{k}-m)\ell_{k,m}^{\dagger},\quad [H_{k},\ell_{k,m}] = -f_{m}(N_{k})\ell_{k,m}, \label{E-29}
\end{align}
where $\mathcal{P}^{(k)}_{iso} = \sum\limits_{n=0}^{\infty}|\psi_{n}^{(k)}\rangle\langle\psi_{n}^{(k)}|$ is the projector onto the subspace $\mathcal{H}^{(k)}_{iso}$ generated by the eigenstates of $H_{k}$ associated to the isospectral part of $H_{0}$. Let us recall that the extreme states belong to the \textit{kernel} of the operator product $\ell^{\dagger}_{k,m}\ell_{k,m}$, which is given by
\begin{equation} \label{E-30}
\ell^{\dagger}_{k,m}\ell_{k,m}=\prod_{l=0}^{m-1}[E(N_{k}-l)-E_0]\prod_{i=1}^{k}[E(N_{k}-m)-\epsilon_i][E(N_{k})-\epsilon_i]\mathcal{P}^{(k)}_{iso}.
\end{equation}
Since they are as well eigenstates of $H_{k}$, then there are $k+m$ extreme states. The first $m$ of them are $\left|\psi_{j}^{(k)}\right\rangle$ , $j=0,1,\cdots ,m-1$, whose energies $E_{j}$ are in the isospectral part of $H_{0}$ and $H_{k}$. The iterated action of $\ell^{\dagger}_{k,m}$ on each $|\psi_{j}^{(k)}\rangle$ generates all the states $\left|\psi_{j+nm}^{(k)}\right\rangle$ of the associated ladder, with energies $E_{j+nm}$. The other $k$ extreme states become $\left|\psi_{\epsilon_{i}}^{(k)}\right\rangle, i=1,2,\cdots ,k$, where each energy $\epsilon_i$ constitutes by itself a single-step ladder, due to $\ell^{\dagger}_{k,m}|\left|\psi_{\epsilon_{i}}^{(k)}\right\rangle=0$. In Fig. \ref{F-2} it is represented the spectrum of $H_{k}$, which is divided into $m$ infinite ladders labeled by the index $j$, and the $k$ single-step ladders.

\section{Coherent states}
The coherent states for the intrinsic algebra of $H_{0}$ and the natural algebra of $H_{k}$ have been generated previously \cite{Veronique2007}. Let us derive now the coherent states for the multiphoton algebra of $H_{0}$ and $H_{k}$, which will be called multiphoton coherent states (MCS).

\subsection{MCS of $\mathbf{H_{0}}$}

Let us look for the MCS $|z\rangle$ of $H_{0}$ as eigenstates of the multiphoton annihilation operator of Eq.~\eqref{E-19}. Since the state space is divided into $m$ orthogonal subspaces, such that their direct sum is the original space, then the coherent states $|z\rangle$ in general will be linear combinations of $m$ orthogonal coherent states (CS) $|z\rangle_{m,\, j}$, where each $|z\rangle_{m,\, j}$ belongs just to the $j$th subspace. Thus, let us restrict ourselves to look for the coherent states $|z\rangle_{m,\, j}$ as eigenstates of the multiphoton annihilation operator \cite{Barut1971}, i.e., 
\begin{equation} \label{E-31}
a_{m}|z\rangle_{m,\, j} = z|z\rangle_{m,\, j},\quad z\in\mathbb{C},\quad j=0,\dots,m-1.
\end{equation}  
Since $|z\rangle_{m,\, j}$ is a linear combination of the states $|\psi_{j+nm}^{(0)}\rangle$, a straightforward calculation leads to 
\begin{equation} \label{E-32}
|z\rangle_{m,\, j} = \frac{1}{\sqrt{\sum\limits_{n=0}^{\infty}\frac{|z|^{2n}}{\rho_{n}^{j}}}}\sum_{n=0}^{\infty}e^{-i\alpha(E_{j+nm}-E_{j})}\frac{z^{n}}{\sqrt{\rho_{n}^{j}}}|\psi_{j+nm}^{(0)}\rangle,\quad j=0,\dots,m-1,
\end{equation}
where 
\begin{equation}\label{E-33}
\rho_{n}^{j} = 
\begin{cases}
\prod\limits_{l=0}^{nm-1}(E_{j+nm-l}-E_{0}) & \mathrm{for}\quad n>0, \\
1 & \mathrm{for}\quad n=0.
\end{cases}
\end{equation}  
It is worth noticing that the coherent state $|z\rangle_{m,\, j}$ reduces to the extreme state $|\psi_{j}^{(0)}\rangle$ for $z=0$.  

A characteristic feature of the standard harmonic oscillator coherent states is that they minimize the Heisenberg uncertainty relation for the standard position and momentum operators. However, in general, it is difficult to calculate algebraically in a closed form the action of such operators on the CS of Eq.~\eqref{E-32}. Moreover, trying to express the standard position and momentum operators in terms of the intrinsic or multiphoton annihilation and creation operators of Eqs.~\eqref{E-14} and \eqref{E-19} is a non-trivial task that goes beyond the scope of this paper. Thus, we shall calculate the uncertainty relation for the naturally associated quadratures, i.e., we will define two operators analogous to the standard position and momentum operators, which will be expressed either in terms of $a$ and $a^{\dagger}$ or $a_{m}$ and $a_{m}^{\dagger}$. Consequently, we will be able to calculate algebraically such uncertainty product. Let us explore first the quadratures defined by
\begin{equation} \label{E-34}
X = \frac{1}{\sqrt{2}}(a^{\dagger} + a),\quad P = \frac{i}{\sqrt{2}}(a^{\dagger} - a).
\end{equation} 
By calculating their corresponding uncertainty relation on the multiphoton coherent state $|z\rangle_{m,\, j}$, which in general is not an eigenstate of $a$, we get that
\begin{equation} \label{E-35}
\left(\Delta X\right)\left(\Delta P\right) = 
\begin{cases}
|a|z\rangle_{m,\, j}|^{2} + \frac{1}{2}\langle[a,a^{\dagger}]\rangle_{m,\, j} & \mathrm{for}\quad m\geq 3,\\
\sqrt{(|a|z\rangle_{m,\, j}|^{2} + \frac{1}{2}\langle[a,a^{\dagger}]\rangle_{m,\, j})^{2} - (\mathrm{Re}(z))^{2}} & \mathrm{for}\quad m=2,\\
\frac{1}{2}\langle[a,a^{\dagger}]\rangle_{1,\ 0} & \mathrm{for}\quad m=1,
\end{cases}
\end{equation}
with
\begin{equation} \label{E-36}
\begin{aligned}
&|a|z\rangle_{m,\, j}|^{2} = \langle a^{\dagger}a\rangle_{m,\, j} =  \frac{1}{\sum\limits_{n=0}^{\infty}\frac{|z|^{2n}}{\rho_{n}^{j}}}\sum\limits_{n=0}^{\infty}\frac{E_{j+nm}-E_{0}}{\rho_{n}^{j}}|z|^{2n},\\
&\langle[a,a^{\dagger}]\rangle_{m,\, j} = \frac{1}{\sum\limits_{n=0}^{\infty}\frac{|z|^{2n}}{\rho_{n}^{j}}}\sum\limits_{n=0}^{\infty}\frac{E_{j+nm+1}-E_{j+nm}}{\rho_{n}^{j}}|z|^{2n}. 
\end{aligned}
\end{equation}
We must mention that for the harmonic oscillator Hamiltonian, the operators $X$ and $P$ in Eq.~\eqref{E-34} are the standard position and momentum operators. However, due to the intrinsic algebra, in general any one-dimensional Hamiltonian with infinite discrete spectrum is factorized by the operators $a$ and $a^{\dagger}$, in the way, $H_{0} = a^{\dagger}a + E_{0}$ \cite{Veronique2007}. Thus, the operator $P$ can be seen as the momentum operator but the operator $X$ just fulfils $X^{2}/2 = V(x) - E_{0} + f(N)$. Therefore, the uncertainty relation for $X$ and $P$ does not supply any new information about the physical position and momentum operators. The other case to be studied arises when defining the operators
\begin{equation} \label{E-37}
X_{m} = \frac{1}{\sqrt{2}}(a_{m}^{\dagger} + a_{m}),\quad P_{m} = \frac{i}{\sqrt{2}}(a_{m}^{\dagger} - a_{m}).
\end{equation}
Since the multiphoton coherent states $|z\rangle_{m,\, j}$ are eigenstates of $a_{m}$ it turns out that
\begin{equation} \label{E-38}
\left(\Delta X_{m}\right)\left(\Delta P_{m}\right) = \frac{\langle a_{m}a_{m}^{\dagger}\rangle_{m,\, j} - |z|^{2}}{2},
\end{equation} 
where
\begin{equation} \label{E-39}
\langle a_{m}a_{m}^{\dagger}\rangle_{m,\, j} = \frac{1}{\sum\limits_{n=0}^{\infty}\frac{|z|^{2n}}{\rho_{n}^{j}}}\sum\limits_{n=0}^{\infty}\frac{|z|^{2n}}{\rho_{n}^{j}}\prod_{l=0}^{m-1}(E_{j+(n+1)m-l}-E_{0}).
\end{equation}
It should be noted that the operators $X_{m}$ and $P_{m}$ in Eq.~\eqref{E-37} can be rewritten as polynomial functions of the operators $X$ and $P$, thus their uncertainty product \eqref{E-38} does not supply further physical information about the system. 

\subsection{MCS of $\mathbf{H_{k}}$}

Similarly, let us look for the MCS $|w\rangle_{m,\, j}$ as eigenstates of the multiphoton annihilation operator $\ell_{k,m}$ of Eq.~\eqref{E-24}, namely,
\begin{equation} \label{E-40}
\ell_{k,m}|w\rangle_{m,\, j} = w|w\rangle_{m,\, j},\quad w\in\mathbb{C}.
\end{equation} 
They are linear combinations of the states $\left|\psi_{n}^{(k)}\right\rangle$ but the isolated states $\left|\psi_{\epsilon_{i}}^{(k)}\right\rangle$ do not contribute to these superpositions since they are in the {\it kernel} of the operator $B_{k}$. Thus, it turns out that 
\begin{equation} \label{E-41}
|w\rangle_{m,\, j} = \frac{1}{\sqrt{\sum\limits_{n=0}^{\infty}\frac{|w|^{2n}}{\varphi_{n}^{j}}}}\sum\limits_{n=0}^{\infty}e^{-i\alpha(E_{j+nm}-E_{j})}\frac{w^{n}}{\sqrt{\varphi_{n}^{j}}}\left|\psi_{j+nm}^{(k)}\right\rangle,
\end{equation} 
with $\varphi_{n}^{j}$ being given by 
\begin{equation} \label{E-42}
\varphi_{n}^{j} = 
\begin{cases}
\frac{\prod_{i=1}^{k}\prod_{p=0}^{n}(E_{j+pm}-\epsilon_{i})^{2}\prod_{l=0}^{nm-1}(E_{j+nm-l}-E_{0})}{\prod_{i=1}^{k}(E_{j+nm}-\epsilon_{i})(E_{j}-\epsilon_{i})} & \mathrm{for}\quad n>0,\\
1 & \mathrm{for}\quad n=0.
\end{cases}
\end{equation}
Once again, for $w=0$ the MCS become the extreme states $\left|\psi_{j}^{(k)}\right\rangle$. It is worth mentioning that, using Eq.~\eqref{E-7}, the MCS $|w\rangle_{m,\, j}$ can be written as
\begin{equation}\label{wz}
	|w\rangle_{m,\, j} = B_{k}^{\dagger}\sqrt{\frac{\prod\limits_{i=1}^{k}\left(E_{j}-\epsilon_{i}\right)}{\sum\limits_{n=0}^{\infty}\frac{|w|^{2n}}{\phi_{n}^{j}}}}\sum\limits_{n=0}^{\infty}\frac{1}{\sqrt{\prod\limits_{p=0}^{n}\prod\limits_{i=1}^{k}\left(E_{j+pm}-\epsilon_{i}\right)^{2}}}e^{-i\alpha\left(E_{j+nm}-E_{j}\right)}\frac{w^{n}}{\sqrt{\rho_{n}^{j}}}|\psi_{j+nm}^{(0)}\rangle.
\end{equation}
Despite the operators $\ell_{k,m}$ and $a_{m}$ are related through Eq.~\eqref{E-24}, one can realize that the linear combination of states $|\psi_{j+nm}^{(0)}\rangle$ in Eq.~\eqref{wz} contains different coefficients as compared with the linear combination of Eq.~\eqref{E-32}, which at first sight cannot be seen as the action of a linear operator acting on the last. Hence, apparently the MCS $|w\rangle_{m,\, j}$ and $|z\rangle_{m,\, j}$ cannot be straightforwardly related to each other.

The quadratures approach allows us to calculate algebraically the uncertainty relation for the operators $X_{k}$ and $P_{k}$ defined in terms of $\ell_{k,1} = \ell_{k}$ and $\ell_{k,1}^{\dagger} = \ell_{k}^{\dagger}$ as follows
\begin{equation} \label{E-43}
X_{k} = \frac{1}{\sqrt{2}}(\ell_{k}^{\dagger} + \ell_{k}),\quad P_{k} = \frac{i}{\sqrt{2}}(\ell_{k}^{\dagger} - \ell_{k}).
\end{equation}
Thus, for the MCS $|w\rangle_{m,\, j}$ which are not necessarily eigenstates of $\ell_{k}$, the uncertainty relation becomes analogous to the one for the quadratures $X$ and $P$, but now it is better to express the result in terms of the products $\ell_{k}^{\dagger}\ell_{k}$ and $\ell_{k}\ell_{k}^{\dagger}$ instead of the commutator $[\ell_{k},\ell_{k}^{\dagger}]$, namely, 
\begin{equation} \label{E-44}
(\Delta X_{k})(\Delta P_{k})=
\begin{cases}
\frac{\langle\ell_{k}^{\dagger}\ell_{k}\rangle_{m,\, j}+\langle\ell_{k}\ell_{k}^{\dagger}\rangle_{m,\, j}}{2} & \mathrm{for}\quad m\geq 3,
\\
\sqrt{\frac{(\langle\ell_{k}^{\dagger}\ell_{k}\rangle_{m,\, j}+\langle\ell_{k}\ell_{k}^{\dagger}\rangle_{m,\, j})^2}{4}-(\text{Re}(w))^2} & \mathrm{for}\quad m=2,
\\
\frac{\langle\ell_{k}\ell_{k}^{\dagger}\rangle_{m,\, j}-|w|^2}{2} & \mathrm{for}\quad m=1,
\end{cases}
\end{equation} 
where
\begin{equation} \label{E-45}
\begin{aligned}
&\langle\ell_{k}^{\dagger}\ell_{k}\rangle_{m,\, j}=|\ell_{k}|w\rangle_{m,\, j}|^2=\frac{\sum\limits_{n=0}^{\infty}\frac{(E_{j+nm}-E_0)\prod\limits_{i=1}^{k}(E_{j+nm}-\epsilon_i)(E_{j+nm-1}-\epsilon_i)}{\varphi_{n}^{j}}|w|^{2n}}{\sum\limits_{n=0}^{\infty}\frac{|w|^{2n}}{\varphi_{n}^{j}}},
\\
&\langle\ell_{k}\ell_{k}^{\dagger}\rangle_{m,\, j}=|\ell_{k}^{\dagger}|w\rangle_{m,\, j}|^2=\frac{\sum\limits_{n=0}^{\infty}\frac{(E_{j+nm+1}-E_0)\prod\limits_{i=1}^{k}(E_{j+nm+1}-\epsilon_i)(E_{j+nm}-\epsilon_i)}{\varphi_{n}^{j}}|w|^{2n}}{\sum\limits_{n=0}^{\infty}\frac{|w|^{2n}}{\varphi_{n}^{j}}}.
\end{aligned}
\end{equation}
Finally, let us define the quadratures $X_{k,m}$ and $P_{k,m}$ in terms of the multiphoton annihilation and creation operators as follows
\begin{equation} \label{E-46}
X_{k,m} = \frac{1}{\sqrt{2}}(\ell_{k,m}^{\dagger}+\ell_{k,m}),\quad P_{k,m} = \frac{i}{\sqrt{2}}(\ell_{k,m}^{\dagger}-\ell_{k,m}).
\end{equation}
Thus, the associated uncertainty relation on the MCS $|w\rangle_{m,\, j}$, which is an eigenstate of $\ell_{k,m}$, becomes
\begin{equation} \label{E-47}
(\Delta X_{k,m})(\Delta P_{k,m})=\frac{\langle\ell_{k,m}\ell_{k,m}^{\dagger}\rangle_{m,\, j}-|w|^2}{2},
\end{equation} 
with
\begin{equation} \label{E-48}
\langle\ell_{k,m}\ell_{k,m}^{\dagger}\rangle_{m,\, j}=|\ell_{k,m}^{\dagger}|w\rangle_{m,\, j}|^2=\frac{\sum\limits_{n=0}^{\infty}\frac{\prod\limits_{l=0}^{m-1}(E_{j+(n+1)m-l}-E_0)\prod\limits_{i=1}^{k}(E_{j+(n+1)m}-\epsilon_i)(E_{j+nm}-\epsilon_i)}{\varphi_{n}^{j}
}|w|^{2n}}{\sum\limits_{n=0}^{\infty}\frac{|w|^{2n}}{\varphi_{n}^{j}}}.
\end{equation}

Note that an intrinsic algebra can be also defined for the Hamiltonian $H_{k}$. Thus, the operators $\ell_{k,m}$ and $\ell_{k,m}^{\dagger}$ can be seen as polynomial deformations of the corresponding intrinsic operators. Consequently, in general the uncertainty relations for the operators $X_{k,m}$ and $P_{k,m}$ of Eq.~\eqref{E-46}, and for the particular case with $m=1$ of Eq.~\eqref{E-43}, do not supply any new information about the physical position and momentum operators of the system.

Since from the very beginning we assumed that the spectrum of the initial Hamiltonian is increasing with respect to $n$, such that $E_{n+1}>E_{n}\ \forall\ n$, and for the SUSY partner potentials the factorization energies are below $E_{0}$, then all resulting infinite series for the uncertainty relations converge. This is true, in particular, for the examples that we will work in the next section, where the energy eigenvalues of $H_{0}$ will be linear or quadratic functions of $n$, i.e., $E(n)\sim n, n^{2}$.

\section{Particular cases}

The theory developed in the previous sections contains the general formulas allowing us to straightforward determine the multiphoton algebra and MCS for most of the known solvable potentials and their $k$th-order SUSY partners. In this work, we will address two examples: the harmonic oscillator and trigonometric P{\"o}schl-Teller potentials. We shall characterize their corresponding algebras and we will illustrate the uncertainty relations for the quadratures associated to the MCS.

\subsection{Harmonic oscillator}  

The harmonic oscillator potential is given by $V_{0}(x) = x^{2}/2$, whose Hamiltonian seen as an energy operator function is linear in $N$, namely, $H_{0} = E(N) = N + 1/2$. When substituting this in Eq.~\eqref{E-21}, one can see that, for example, the energy operator evaluated in $N+m-l$ gives $E(N+m-l)= N+m-l+1/2$ while $E_{0} = E(0) = 1/2$, thus obtaining that the multiphoton algebra looks like
\begin{equation} \label{E-49}
\begin{aligned}
&[a_{m},a_{m}^{\dagger}] = \prod_{l=0}^{m-1}\left(H_{0} + m -l - \frac{1}{2}\right) - \prod_{l=0}^{m-1}\left(H_{0} - l - \frac{1}{2}\right), \\
& [H_{0},a_{m}^{\dagger}] = ma_{m}^{\dagger},\quad [H_{0},a_{m}] = -ma_{m}.
\end{aligned}
\end{equation} 
Meanwhile, on the isospectral subspace $\mathcal{H}^{(k)}_{iso}$ the Hamiltonian becomes $H_{k}= E(N_{k}) = N_{k}+1/2$. Hence, from Eqs.~\eqref{E-28} and \eqref{E-29}, the algebra for the SUSY partners becomes 
\begin{align}
&\begin{aligned}
&[\ell_{k,m},\ell_{k,m}^{\dagger}] = \Bigg\{\prod_{l=0}^{m-1}\left(H_{k} + m - l - \frac{1}{2}\right)\prod_{i=1}^{k}\left(H_{k} + m - \epsilon_{i}\right)\left(H_{k} - \epsilon_{i}\right)  \\
& \qquad\qquad\ \ -\prod_{l=0}^{m-1}\left(H_{k} - l - \frac{1}{2}\right)\prod_{i=1}^{k}\left(H_{k} - m - \epsilon_{i}\right)\left(H_{k} - \epsilon_{i}\right)\Bigg\}\mathcal{P}^{(k)}_{iso},
\end{aligned}\label{E-50}\\
& [H_{k},\ell_{k,m}^{\dagger}] = m\ell_{k,m}^{\dagger},\quad [H_{k},\ell_{k,m}] = -m\ell_{k,m}. \label{E-51}
\end{align}
Note that in both algebras the commutator between the annihilation and creation operators is a polynomial in the corresponding Hamiltonian in the subspace associated to the initial spectrum, which for $H_{0}$ coincides with the full Hilbert space. Furthermore, the commutator between the Hamiltonian and a ladder operator is proportional to the same ladder operator, the factor being the difference between two consecutive energy levels in the same ladder, which here is a constant (see Fig. \ref{F-2}).

\begin{figure}[t]
\subfigure[]{\includegraphics[scale=0.41]{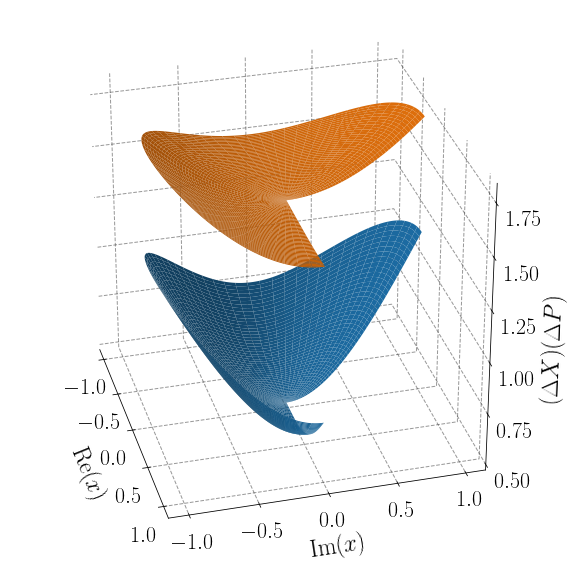}}
\subfigure[]{\includegraphics[scale=0.41]{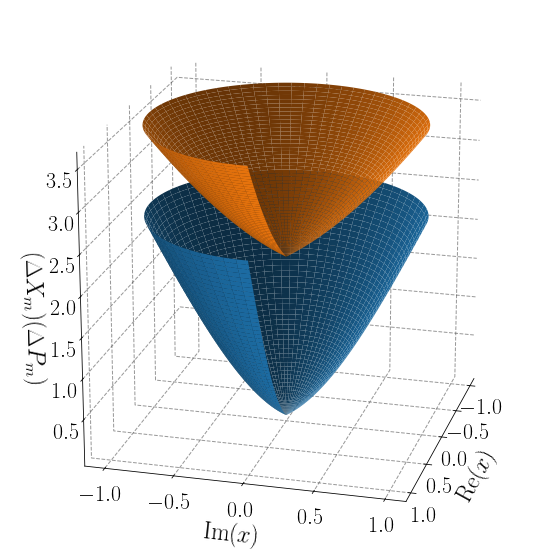}}
\caption{(a) Uncertainty relations for the ``even" and ``odd" CS of the harmonic oscillator potential associated to the quadratures of the intrinsic annihilation and creation operators, where the minimum values are $1/2$ for the ``even" state and $3/2$ for the ``odd" one. (b) The corresponding quantities for the multiphoton operators are shown, and the minimum values are now $0$ for the ``even" state and $2$ for the ``odd" one.}\label{F-3}
\end{figure}

\begin{figure}[t]
\subfigure[]{\includegraphics[scale=0.42]{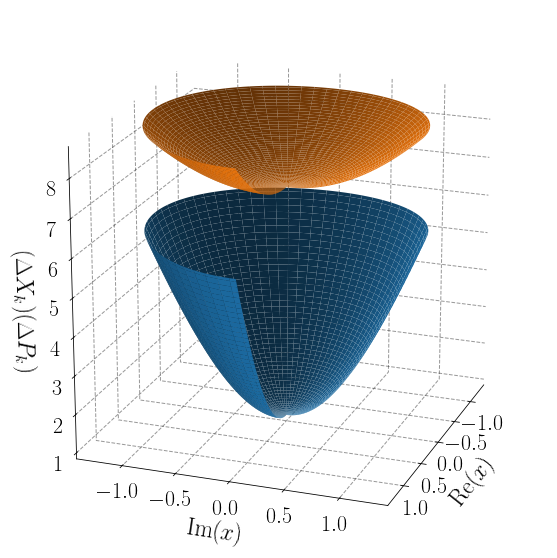}}
\subfigure[]{\includegraphics[scale=0.42]{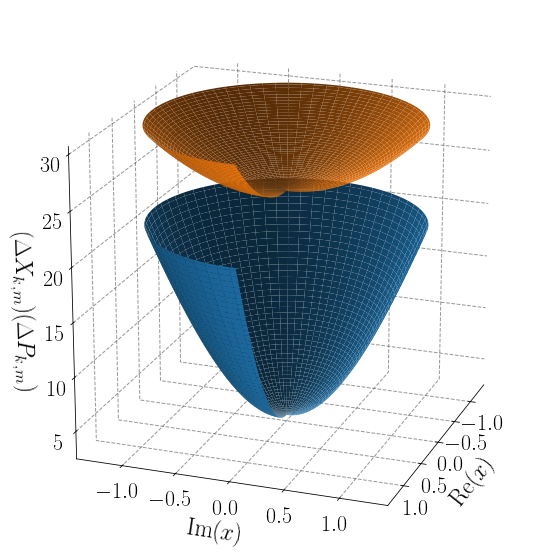}}
\caption{(a) Uncertainty relations for the ``even" and ``odd" CS of the first-order harmonic oscillator SUSY partner potential with factorization energy $\epsilon = -1/2$, for the quadratures associated to the annihilation and creation operators $\ell_{k}$, $\ell^{\dagger}_{k}$. The minimum values achieved are $1$ for the ``even" state and $7$ for the ``odd" one. (b) The multiphoton operators $\ell_{k,m}$, $\ell_{k,m}^{\dagger}$ are now used, and the minimum values obtained are $3$ for the ``even" state and $24$ for the ``odd" one.} \label{F-4}
\end{figure}

On the other hand, from Eqs.~\eqref{E-32} and \eqref{E-33}, the coherent states $|z\rangle_{m,\, j}$ for the oscillator potential $V_{0}(x)$ turn out to be 
\begin{equation} \label{MCSHO}
	|z\rangle_{m,\, j} = \frac{1}{\sqrt{\sum\limits_{n=0}^{\infty}\frac{|z|^{2n}}{(j+nm)!}}}\sum\limits_{n=0}^{\infty}e^{-i\alpha nm}\frac{z^{n}}{\sqrt{(j+nm)!}}|\psi_{j+nm}^{(0)}\rangle,\quad j=0,\dots,m-1.
\end{equation}
In particular, with $\alpha = 0$ and $m = 1$ ($j=0$), the previous coherent states reduce to the standard ones
\begin{equation} \label{E-52}
|z\rangle_{1,0} = e^{-\frac{|z|^{2}}{2}}e^{za^{\dagger}}|\psi_{0}^{(0)}\rangle,
\end{equation}  
while for $m = 2$ the so-called ``even" and ``odd" CS are recovered, which are given by 
\begin{equation} \label{E-53}
|z\rangle_{2,0} = \sqrt{\mathrm{sech}|z|}\sum\limits_{n=0}^{\infty}\frac{z^{n}|\psi_{2n}^{(0)}\rangle}{\sqrt{(2n)!}},\quad |z\rangle_{2,1} = \sqrt{|z|\mathrm{csch}|z|}\sum\limits_{n=0}^{\infty}\frac{z^{n}|\psi_{2n+1}^{(0)}\rangle}{\sqrt{(2n+1)!}}.
\end{equation}
For the SUSY partners, from Eqs.~\eqref{E-41} and \eqref{E-42}, the MCS can be written as follows
\begin{equation}\label{MCSHOk}
|w\rangle_{m,\, j} = \frac{1}{\sqrt{\sum\limits_{n=0}^{\infty}\frac{|w|^{2n}}{(j+nm)!}\prod\limits_{i=1}^{k}\frac{(j+nm+1/2-\epsilon_{i})}{[(j+nm+1/2-\epsilon_{i})!]^{2}}}}\sum\limits_{n=0}^{\infty}\frac{e^{-i\alpha nm}w^{n}|\psi_{j+nm}^{(k)}\rangle}{\sqrt{(j+nm)!\prod\limits_{i=1}^{k}\frac{[(j+nm+1/2-\epsilon_{i})!]^{2}}{(j+nm+1/2- \epsilon_{i})}}},
\end{equation}
where $j=0,\dots,m-1$. In particular, for the first-order SUSY partner potential, i.e., taking $k = 1$ and the factorization energy as $\epsilon_{1} = -1/2$, the ``even" and ``odd" CS, with $m=2$, are given by
\begin{equation}\label{E-54}
|w\rangle_{2,0} = \frac{\sum\limits_{n=0}^{\infty}\sqrt{\frac{(2n+1)}{(2n)![(2n+1)!]^{2}}}w^{n}|\psi_{2n}^{(1)}\rangle}{\sqrt{\sum\limits_{n=0}^{\infty}\frac{(2n+1)|w|^{2n}}{(2n)![(2n+1)!]^{2}}}},\quad |w\rangle_{2,1} = \frac{\sum\limits_{n=0}^{\infty}\sqrt{\frac{(2n+2)}{(2n+1)![(2n+2)!]^{2}}}w^{n}|\psi_{2n+1}^{(1)}\rangle}{\sqrt{\sum\limits_{n=0}^{\infty}\frac{(2n+2)|w|^{2n}}{(2n+1)![(2n+2)!]^{2}}}}.
\end{equation}
The uncertainty relations for the quadratures arising in Eqs.~\eqref{E-35} and \eqref{E-38}, associated to the ``even" and ``odd" CS of the harmonic oscillator potential given in Eq.~\eqref{E-53}, are shown in Fig. \ref{F-3}. In Fig. \ref{F-4} it is plotted the uncertainty relations of Eqs.~\eqref{E-44} and \eqref{E-47} for the CS given in Eq.~\eqref{E-54} associated to the first-order SUSY partner potentials of the harmonic oscillator with factorization energy $\epsilon_{1} = -1/2$.     

\subsection{Trigonometric P{\"o}schl-Teller potential}

The initial potential is now $V_{0}(x) = \nu(\nu - 1)/(2\cos^{2} x)$, $\nu > 1$. The energy operator in terms of the number operator becomes $H_{0} = E(N) = (N + \nu)^{2}/2$, thus the commutators defining the multiphoton algebra of $H_{0}$, from Eq.~\eqref{E-21}, turn out to be
\begin{equation} \label{E-55}
\begin{aligned}
&[a_{m},a_{m}^{\dagger}] = \textstyle\prod\limits_{l=0}^{m-1}\left(H_{0}-(m-l)(N+l-m)-\frac{(m-l+\nu)^{2}}{2}\right) - \prod\limits_{l=0}^{m-1}\left(H_{0}-l(N-l)-\frac{(l+\nu)^{2}}{2}\right), \\
&[H_{0},a_{m}^{\dagger}] = \frac{m}{2}[2(N + \nu)- m]a_{m}^{\dagger},\quad [H_{0},a_{m}] = -\frac{m}{2}[2(N + \nu)+ m]a_{m}.
\end{aligned}
\end{equation}
On the other hand, in the isospectral subspace the energy operator is $H_{k}=E(N_{k}) = (N_{k}+ \nu)^{2}/2$; when substituting this in Eqs.~\eqref{E-28} and \eqref{E-29} the algebra of the SUSY partner Hamiltonian $H_{k}$ is characterized by
\begin{align} 
&\begin{aligned}
\textstyle[\ell_{k,m},\ell_{k,m}^{\dagger}] &= \Bigg\{\prod\limits_{l=0}^{m-1}\left[H_{k} - (m-l)(N_{k}+l-m) - \frac{(m-l+\nu)^{2}}{2}\right]\\
&\textstyle\times \prod\limits_{i=1}^{k}\Big[H_{k} + \frac{m}{2}(m+2(N_{k}+\nu))-\epsilon_{i}\Big]\left(H_{k} - \epsilon_{i}\right)- \prod\limits_{l=0}^{m-1}\left[H_{k}-l(N_{k}-l)-\frac{(l+\nu)^{2}}{2}\right]\\
&\textstyle \times\prod\limits_{i=1}^{k}\Big[H_{k}+\frac{m}{2}(m-2(N_{k}+\nu))-\epsilon_{i}\Big]\left(H_{k}-\epsilon_{i}\right)\Bigg\}\mathcal{P}^{(k)}_{iso},
\end{aligned} \label{E-56}\\
&[H_{k},\ell_{k,m}^{\dagger}] = \frac{m}{2}[2(N_{k} + \nu)- m]\ell_{k,m}^{\dagger},\quad  [H_{k},\ell_{k,m}] = -\frac{m}{2}[2(N_{k} + \nu)+ m]\ell_{k,m}. \label{E-57}
\end{align}
Note that the algebras of Eqs.~(\ref{E-55},~\ref{E-56},~\ref{E-57}) for the trigonometric P{\"o}schl-Teller potential and its $k$th-order SUSY partners have non-constant factors in the commutators of the Hamiltonian with the corresponding ladder operators, thus defining non-constant energy gaps between two consecutive energy steps in the ladders (see Fig. \ref{F-2}). However, the commutators between the two ladder operators is a polynomial function in the corresponding Hamiltonians in the subspace associated to the initial spectrum. 

\begin{figure}[t]
\subfigure[]{\includegraphics[scale=0.41]{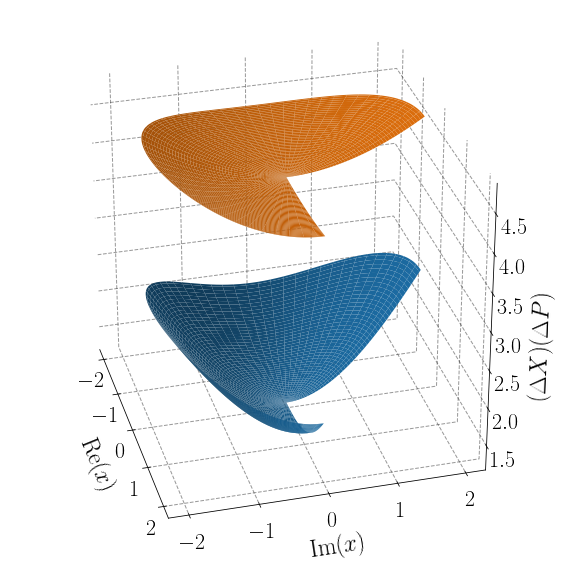}}
\subfigure[]{\includegraphics[scale=0.41]{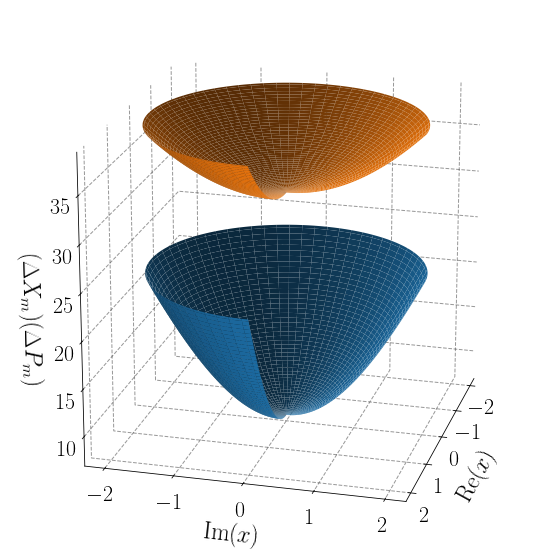}}
\caption{(a) Uncertainty relations for the CS of the trigonometric P{\"o}schl-Teller potential for the quadratures of the intrinsic annihilation and creation operators. The minimum values achieved are $5/4$ for $|z\rangle_{2,0}$ and $17/4$ for $|z\rangle_{2,1}$. (b) The corresponding multiphoton operators are now used, and the minimum values obtained are $15/2$ for $|z\rangle_{2,0}$ and $63/2$ for $|z\rangle_{2,1}$. The potential parameter taken is $\nu = 2$.}\label{F-5}
\end{figure}

\begin{figure}[t]
\begin{center}
\subfigure[]{\includegraphics[scale=0.41]{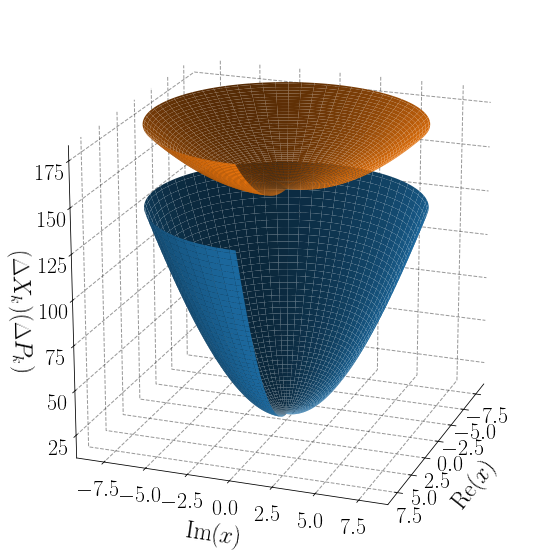}}
\subfigure[]{\includegraphics[scale=0.41]{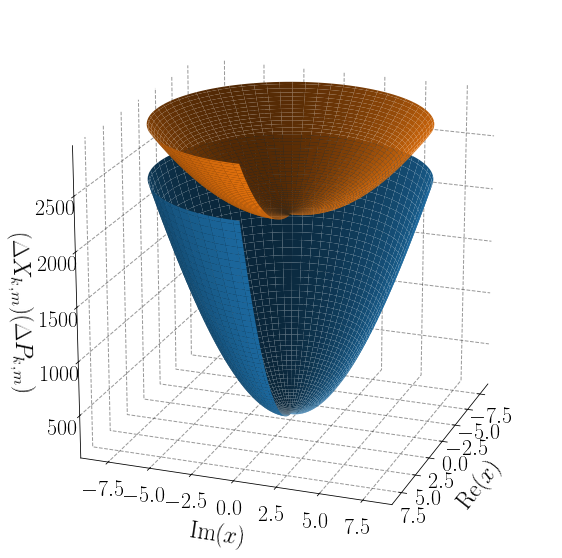}}
\caption{(a) Uncertainty relations for the CS of the first-order SUSY partner of the trigonometric P{\"o}schl-Teller potential with $\epsilon = -1/2$, for the quadratures of the annihilation and creation operators $\ell_{k}$, $\ell^{\dagger}_{k}$ . The minimum values achieved are $125/8$ for $|w\rangle_{2,0}$ and $1145/8$ for $|w\rangle_{2,1}$. (b) The corresponding multiphoton operators are now used, and the minimum values obtained are $1275/8$ for $|w\rangle_{2,0}$ and $4095/2$ for $|w\rangle_{2,1}$. The potential parameter taken is $\nu = 2$.} \label{F-6}
\end{center}
\end{figure}

On the other hand, from Eqs.~\eqref{E-32} and \eqref{E-33}, the MCS for the trigonometric P{\"o}schl-Teller potential $|z\rangle_{m,\, j}$ are given by 
\begin{equation}\label{MCSPT}
	|z\rangle_{m,\, j} = \frac{1}{\sqrt{\sum\limits_{n=0}^{\infty}\frac{2^{nm}|z|^{2n}}{(j+nm)!(j+nm+2\nu)!}}}\sum\limits_{n=0}^{\infty}\frac{2^{nm/2}e^{-i\frac{\alpha}{2}nm[nm+2(j+\nu)]}z^{n}}{\sqrt{(j+nm)!(j+nm+2\nu)!}}|\psi_{j+nm}^{(0)}\rangle
\end{equation}
with $j=0,\dots,m-1$. Taking $\alpha = 0$ and $m = 2$, the ``even" and ``odd" CS can be written as
\begin{equation}\label{E-58}
\begin{aligned}
&|z\rangle_{2,0} = \frac{1}{\sqrt{\sum\limits_{n=0}^{\infty}\frac{2^{2n}|z|^{2n}}{(2n)![2(n+\nu)]!}}}\sum\limits_{n=0}^{\infty}\frac{2^{n}z^{n}}{\sqrt{(2n)![2(n+\nu)]!}}|\psi_{2n}^{(0)}\rangle,\\ 
&|z\rangle_{2,1} = \frac{1}{\sqrt{\sum\limits_{n=0}^{\infty}\frac{2^{2n}|z|^{2n}}{(2n+1)![2(n+\nu)+1]!}}}\sum\limits_{n=0}^{\infty}\frac{2^{n}z^{n}}{\sqrt{(2n+1)![2(n+\nu)+1]!}}|\psi_{2n+1}^{(0)}\rangle.
\end{aligned}
\end{equation}   
Meanwhile, from Eqs.~\eqref{E-41} and \eqref{E-42}, the corresponding CS $|w\rangle_{m,\, j}$ for the associated first-order SUSY partner with $k = 1$ and $\epsilon_{1} = -1/2$,  become
\begin{equation} \label{MCSPTk}
\begin{aligned}
|w\rangle_{m,\, j} &= \frac{1}{\sqrt{\sum\limits_{n=0}^{\infty}\frac{2^{nm}|w|^{2n}}{(j+nm)!(j+nm+2\nu)!}\prod\limits_{i=1}^{k}\frac{\left(\frac{(j+nm+\nu)^{2}}{2}-\epsilon_{i}\right)}{\prod\limits_{p=0}^{n}\left(\frac{(j+pm+\nu)^{2}}{2}-\epsilon_{i}\right)^{2}}}}\\
&\times\sum\limits_{n=0}^{\infty}\sqrt{\frac{2^{nm}}{(j+nm)!(j+nm+2\nu)!}\prod\limits_{i=1}^{k}\frac{\left(\frac{(j+nm+\nu)^{2}}{2}-\epsilon_{i}\right)}{\prod\limits_{p=0}^{n}\left(\frac{(j+pm+\nu)^{2}}{2}-\epsilon_{i}\right)^{2}}}\\
&\times e^{-i\frac{\alpha}{2}nm[nm+2(j+\nu)]}w^{n}|\psi_{j+nm}^{(k)}\rangle,
\end{aligned}
\end{equation}
where $j=0,\dots,m-1$. The corresponding ``even" and ``odd" CS, with $m=2$, turn out to be
\begin{equation}\label{E-59}
\begin{aligned}
|w\rangle_{2,0} &= \frac{1}{\sqrt{\sum\limits_{n=0}^{\infty}\frac{2^{4n+1}[(2n+\nu)^{2}+1]|w|^{2n}}{(2n)![2(n+\nu)]!\prod\limits_{p=0}^{n}[(2n+\nu)^{2}+1]^{2}}}}\sum\limits_{n=0}^{\infty}\sqrt{\frac{2^{4n+1}[(2n+\nu)^{2}+1]}{(2n)![2(n+\nu)]!\prod\limits_{p=0}^{n}[(2p+\nu)^{2}+1]^{2}}}w^{n}|\psi_{2n}^{(1)}\rangle,\\
|w\rangle_{2,1} &= \frac{1}{\sqrt{\sum\limits_{n=0}^{\infty}\frac{2^{4n+1}[(2n+1+\nu)^{2}+1]|w|^{2n}}{(2n+1)![2(n+\nu)+1]!\prod\limits_{p=0}^{n}[(2n+1+\nu)^{2}+1]^{2}}}}\\
&\times\sum\limits_{n=0}^{\infty}\sqrt{\frac{2^{4n+1}[(2n+1+\nu)^{2}+1]}{(2n+1)![2(n+\nu)+1]!\prod\limits_{p=0}^{n}[(2p+1+\nu)^{2}+1]^{2}}}w^{n}|\psi_{2n+1}^{(1)}\rangle.
\end{aligned}
\end{equation}  
Fig. \ref{F-5} shows the uncertainty relations associated to the quadratures for the CS of the trigonometric P{\"o}schl-Teller potential given in Eq.~\eqref{E-58}. In Fig. \ref{F-6}, the uncertainty relations for the CS of the associated first-order SUSY partner of Eq.~\eqref{E-59} are shown.

\section{Conclusions}

In this paper, the multiphoton algebras ruling a class of one dimensional Hamiltonians with an infinite discrete spectrum, the corresponding $k$th-order SUSY partners and their associated multiphoton algebras, have been studied. In the examples addressed here, these algebras turn out to be polynomial functions of the corresponding Hamiltonians in the subspace associated to the isospectral part. The energy gap between two adjacent energy levels is constant for the harmonic oscillator potential, while it is linear in the quantum number $n$ for the trigonometric P{\"o}schl-Teller potential. The multiphoton algebra produces, naturally, the division of the spectrum of $H_0$ into $m$ infinite energy ladders, consequently the corresponding Hilbert space is split into $m$ orthogonal subspaces whose direct sum is the original Hilbert space. It is worth noting that the spectrum of the $k$th-order SUSY partners $H_{k}$ have $k$ extra single-step ladders, additional to the spectrum of the initial Hamiltonian $H_{0}$. The generated multiphoton coherent states for the harmonic oscillator potential reduce to the standard ones and to the ``even" and ``odd" states for the right choice of the $m$th power of the ladders operators. Finally, it would be interesting to evaluate the uncertainty relations for the physical position and momentum operators of our system, a task that will require extensive numerical calculations, which are outside the scope of this paper. 

\section*{Acknowledgments}
This work was supported by CONACYT(Mexico), project FORDECYT-PRONACES/61533/2020. The authors thank to the referees for useful comments and suggestions that helped to improve this paper.

\bibliography{multiphoton_algebras_biblio}
\bibliographystyle{unsrt}
\addcontentsline{toc}{section}{References} 

\end{document}